\documentclass[twocolumn,amsmath,amssymb,superscriptaddress,notitlepage]{revtex4-1}
\usepackage{graphics}
\usepackage{bm}
\usepackage{dcolumn}
\usepackage{natbib}
\usepackage{epstopdf}
\usepackage{float}
\usepackage{graphicx}
\usepackage{epsfig}
\usepackage[pdfstartview=FitH]{hyperref}
\usepackage{color}
\usepackage{appendix}
\usepackage{ulem}

\newcommand{\kk}{{\bf k}}

\newcommand{\rr}{{\bf r}}

\begin{document}
\title{Quantum Hall effects of exciton condensate in topological flat bands}
\author{Tian-Sheng Zeng}
\affiliation{School of Science, Westlake University, Hangzhou 310024, China and \\
Institute of Natural Sciences, Westlake Institute for Advanced Study, Hangzhou 310024, China}
\author{D. N. Sheng}
\affiliation{Department of Physics and Astronomy, California State University, Northridge, California 91330, USA}
\author{W. Zhu}
\affiliation{School of Science, Westlake University, Hangzhou 310024, China and \\
Institute of Natural Sciences, Westlake Institute for Advanced Study, Hangzhou 310024, China}
\date{\today}
\begin{abstract}
Tunable exciton condensates in two dimensional electron gas systems under strong magnetic field exhibits anomalous Hall transport owing to mutual Coulomb coupling, and have attracted a lot of research activity. Here, we explore another framework using topological flat band models in the absence of Landau levels, for realizing the many-body exciton phases of two-component fermions under strong intercomponent interactions. By developing new diagnosis based on the state-of-the-art density-matrix renormalization group and exact diagonalization, we show the theoretical discovery of the emergence of Halperin (111) quantum Hall effect at a total filling factor $\nu=1$ in the lowest Chern band under strong Hubbard repulsion, which is classified by the unique ground state with bulk charge insulation and spin superfluidity, The topological nature is further characterized by one edge branch of chiral propagating Luttinger modes with level counting $1,1,2,3,5,7$ in consistent with the conformal field theory description. Moreover, with nearest-neighbor repulsions, we propose the Halperin (333) fractional quantum Hall effect at a total filling factor $\nu=1/3$ in the lowest Chern band.
\end{abstract}
\maketitle

\section{Introduction}

As extensions of Laughlin wavefunction, the intercomponent correlation of two-component quantum Hall effects was proposed as Halperin's two-component $(mmn)$ wavefunction in Ref.~\cite{Halperin1983}, described by the $\mathbf{K}=\begin{pmatrix}
m & n\\
n & m\\
\end{pmatrix}$ matrix within the framework of the field theory~\cite{Wen1992a,Wen1992b,Blok1990a,Blok1990b,Blok1991}. Two-component quantum Hall effects, where the phrase ``two-component'' represents a generic label for spin or pseudo-spin (bilayer, double well, etc.) quantum number, could exhibit tremendously richer physics than single-component systems, attributing to intercomponent interaction in two-component systems. Experimentally, such exotic intercomponent correlation for two-component integer and fractional quantum Hall states at $\nu=1$ and $\nu=1/2$ in the lowest Landau level had been confirmed~\cite{Eisenstein1992,Suen1992}, in consistency with several theoretical proposals of Halperin's (111) exciton superfluid~\cite{Wen1992,Moon1995,Yang1996} and (331) fractional quantum Hall state~\cite{Yoshioka1988,Yoshioka1989,He1991,He1993}.

To gain a better understanding of the internal structure of multicomponent quantum Hall states, it is highly desirable to investigate the diagonal and off-diagonal elements of the $\mathbf{K}$ matrix which describe intracomponent and intercomponent Chern-Simons gauge-field couplings respectively, and can be derived from the inverse of the Chern number matrix for gapped quantum Hall states~\cite{Zeng2017,Zeng2018,Zeng2019,Zeng2020}. However a peculiar property of Halperin $(mmm)$ quantum Hall states is that they host intercomponent tunneling and counterflow transport anomalies in Hall resistance measurements~\cite{Eisenstein2014}, due to the superfluidity of exciton condensate in which particles in one component are coupled to holes in the other component. Such coherent exciton transport, which serves as a hallmark signature of Halperin (111) quantum Hall effect, has inspired a series of motivating quantum Hall drag measurements in different setups~\cite{Kellogg2002,Tutuc2004,Finck2010,Nandi2012,Liu2017,Li2017,Eisenstein2019}. The intercomponent Coulomb drag transport in most recent experiments also serves as a primary proof in searching new types of correlated many-body topological states in two parallel graphene layers~\cite{Li2019,Liu2019}.

Recently the arise of topological flat band models has fostered an excitingly new platform for studying the quantum Hall effect without the conventional Landau level~\cite{Sun2011,Neupert2011,Sheng2011,Tang2011,Wang2011,Regnault2011} (See the recent reviews in Refs.\cite{BL2013,Cayssol2013} for extended literatures). On the experimental side, realization of topological Haldane-honeycomb band provides a highly tunable system to explore intercomponent correlated states of two-component quantum fermionic $^{40}$K gases with strong Hubbard repulsion~\cite{Jotzu2014}, and the two-component correlated charge pumping can be implemented using spin- and density-resolved microscopy~\cite{Koepsell2020}. Also, various types of topological bands are proposed in multi-layer heterostructure, such as moir\'e flat bands in twisted multilayer graphene~\cite{Zhang2019,Lee2019,Dai2019,Chen2019} with tunable correlated ferromagnetism observed~\cite{Chen2020}, and some of fractionalised interacting phases in such topological bands dubbed ``fractional Chern insulators'' have been experimentally observed~\cite{Spanton2017}. These related experimental advances, thus enabling the control and study of multi-layered systems, would open up new relevant prospects across a broader class of two-component Halperin $(mmm)$ quantum Hall effects for interacting fermions in topological lattice models, where a compelling theoretical evidence of them is still lacking and demanding which is the focus of our work.

In this work, we theoretically propose two-component Halperin $(mmm)$ quantum Hall effects emerging in topological flat bands through the state-of-the-art density-matrix renormalization group (DMRG) and exact diagonalization (ED) simulations with strong interactions in the thermodynamic limit, and elucidate the interaction controlling of the topological exciton condensate of two-component systems. We have studied the characteristic topological degeneracy, the excitation gap, the topologically invariant Chern number, charge pumping, off-diagonal long range order and entanglement spectrum of the ground states, which depict the topological information from $\mathbf{K}$ matrix.

This paper is organized as follows. In Sec.~\ref{model}, we give a description of the model Hamiltonian of interacting two-component fermions in different topological lattice models, such as $\pi$-flux checkerboard and Haldane-honeycomb lattices. In Sec.~\ref{ground}, we study the many-body ground states of these two-component fermions in the strong interaction regime, present detailed numerical results of Halperin (111) state at fillings $\nu=1$ in Sec.~\ref{H111}, and further discuss the topological signatures of Halperin (333) state under nearest-neighboring repulsion at fillings $\nu=1/3$ in Sec.~\ref{H333}. Finally, in Sec.~\ref{summary}, we summarize our results and discuss the prospect of investigating topological exciton condensate in two-component systems.

\section{Model and Method}\label{model}

Here, we start from the following Hamiltonian of interacting spinfull fermions in two typical topological lattice models, such as the $\pi$-flux checkerboard (CB) lattice,
\begin{align}
  H_{CB}=&\!\sum_{\sigma}\!\Big[-t\!\!\sum_{\langle\rr,\rr'\rangle}\!e^{i\phi_{\rr'\rr}}c_{\rr',\sigma}^{\dag}c_{\rr,\sigma}
  -\!\!\!\!\sum_{\langle\langle\rr,\rr'\rangle\rangle}\!\!\!t_{\rr,\rr'}'c_{\rr',\sigma}^{\dag}c_{\rr,\sigma}\nonumber\\
  &-t''\!\!\!\sum_{\langle\langle\langle\rr,\rr'\rangle\rangle\rangle}\!\!\!\! c_{\rr',\sigma}^{\dag}c_{\rr,\sigma}+H.c.\Big]+V_{int},\label{cbl}
\end{align}
and Haldane-honeycomb (HC) lattice
\begin{align}
  H_{HC}=&\!\sum_{\sigma}\!\Big[-t\!\!\sum_{\langle\rr,\rr'\rangle}\!\! c_{\rr',\sigma}^{\dag}c_{\rr,\sigma}-t'\!\!\sum_{\langle\langle\rr,\rr'\rangle\rangle}\!\!e^{i\phi_{\rr'\rr}}c_{\rr',\sigma}^{\dag}c_{\rr,\sigma}\nonumber\\
  &-t''\!\!\sum_{\langle\langle\langle\rr,\rr'\rangle\rangle\rangle}\!\!\!\! c_{\rr',\sigma}^{\dag}c_{\rr,\sigma}+H.c.\Big]+V_{int},\label{hcl}
\end{align}
where $c_{\rr,\sigma}^{\dag}$ is the particle creation operator of spin $\sigma=\uparrow,\downarrow$ at site $\rr$, $\langle\ldots\rangle$,$\langle\langle\ldots\rangle\rangle$ and $\langle\langle\langle\ldots\rangle\rangle\rangle$ denote the nearest-neighbor, the next-nearest-neighbor, and the next-next-nearest-neighbor pairs of sites, respectively. The flat band limit is taken with the tunnel couplings $t'=0.3t,t''=-0.2t,\phi=\pi/4$ for checkerboard lattice~\cite{Zeng2015}, while $t'=0.6t,t''=-0.58t,\phi=2\pi/5$ for honeycomb lattice~\cite{Wang2011}. We take the on-site and nearest-neighbor interactions with SU(2) symmetry,
\begin{align}
  V_{int}=U\sum_{\rr}n_{\rr,\uparrow}n_{\rr,\downarrow}+V\sum_{\sigma,\sigma'}\sum_{\langle\rr,\rr'\rangle}n_{\rr',\sigma'}n_{\rr,\sigma}
\end{align}
where $n_{\rr,\sigma}$ is the particle number operator of spin $\sigma$ at site $\rr$. Here, $U$ is the strength of the onsite interaction while $V$ is the strength of nearest-neighbor interaction.

We perform ED calculations on the many-body ground state of the model Hamiltonian Eqs.~\ref{cbl} and~\ref{hcl} in a finite system of $N_x\times N_y$ unit cells (the total number of sites is $N_s=2\times N_x\times N_y$), up to $N_s=20$. The total filling of the lowest Chern band is $\nu=\nu_{\uparrow}+\nu_{\downarrow}=2(N_{\uparrow}+N_{\downarrow})/N_s$, where $N_{\uparrow},N_{\downarrow}$ are the particle numbers with $U(1)\times U(1)$-symmetry. With the translational symmetry, the energy states are labeled by the total momentum $K=(K_x,K_y)$ in units of $(2\pi/N_x,2\pi/N_y)$ in the Brillouin zone. For larger systems, we exploit both finite and infinite DMRG on the cylindrical geometry. We keep the maximal bond dimension up to $M=8000$ in infinite DMRG, which leads to excellent convergence for the results we report here. In infinite DMRG, the geometry of cylinders is open boundary condition in the $x$ direction and periodic boundary condition in the $y$ direction. Our comprehensive DMRG and ED studies can access large system sizes to establish the emergence of quantum Hall states at filling factors $\nu=1/m$ for two-component fermions (odd $m=1,3$).

\section{The many-body ground states}\label{ground}

\begin{figure}[b]
  \includegraphics[height=1.8in,width=3.36in]{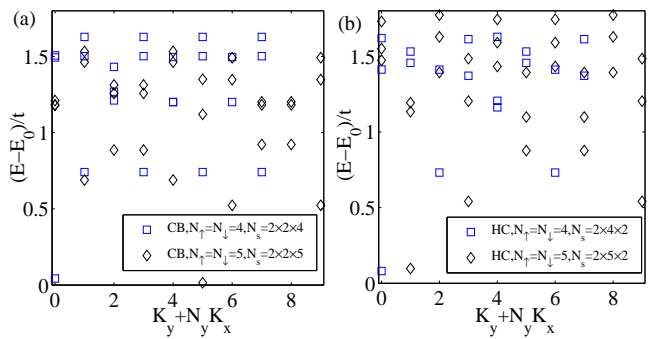}
  \caption{\label{energy} (Color online) Numerical ED results for the low energy spectrum of two-component fermionic systems $\nu=1$ with $U(1)\times U(1)$-symmetry in different topological lattices at $U\gg t,V=0$ for (a) $\pi$-flux checkerboard lattice and (b) Haldane-honeycomb lattice.}
\end{figure}

In this section, we first examine systematically the topological properties of the many-body ground states at $\nu=1,U\gg t,V=0$, which becomes an easy-plane ferromagnet. Having identified the properties of Halperin (111) state which is most favored, we next describe numerical studies of Halperin (333) state at $\nu=1/3,U,V\gg t$.

\subsection{Halperin (111) state}\label{H111}

\begin{figure}[t]
  \includegraphics[height=1.7in,width=3.36in]{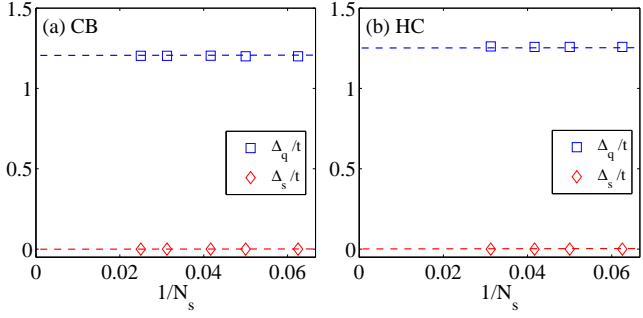}
  \caption{\label{gap} (Color online) Numerical results for the charge-hole gap $\Delta_q$ and spin excitation gap $\Delta_s$ of two-component fermionic systems $\nu=1$ at $U\gg t,V=0$ for (a) $\pi$-flux checkerboard lattice and (b) Haldane-honeycomb lattice. Results are obtained using ED for $N_s=16,20$, and results are obtained using DMRG for $N_s=24,32,40$.}
\end{figure}

For Halperin quantum Hall state characterized by the $\mathbf{K}=\begin{pmatrix}
1 & 1\\
1 & 1\\
\end{pmatrix}$ matrix, the charge channel remain insulating, and the system displays the quantized Hall effect, similar to usual $\nu=1$ integer quantum Hall effect. However the spin channel condenses and the system becomes a gapless superfluid~\cite{Wen1995}. First, we demonstrate the unique ground state degeneracy on periodic lattice under strong onsite Hubbard repulsion $U\gg t,V=0$. In Figs.~\ref{energy}(a) and~\ref{energy}(b), for different topological systems in strong interacting regime, we find that, there exists a well-defined single ground state separated from higher energy levels by a robust gap. In the spin notation, a variational ansatz describing the particle-hole pairs of Halperin (111) state can be written as
\begin{align}
  |\psi\rangle=\prod_{\kk}\frac{1}{\sqrt{2}}(\chi_{\kk,\uparrow}^{\dag}+e^{i\varphi}\chi_{\kk,\downarrow}^{\dag})|0\rangle,\label{wave}
\end{align}
where $\chi_{\kk,\sigma}^{\dag}$ creates a Bloch fermion of spin $\sigma$ and momentum $\kk$ in the lowest Chern band and $\varphi$ is the phase for particle-hole pairs, which can take certain value  without any energy cost. For strong Hubbard repulsion $U\gg t$, each of the single-particle orbitals in the lowest Chern band is occupied with only one particle at $\nu=1$ and the total momentum $K=(\sum_{i=1}^{N} k_x^i,\sum_{i=1}^{N} k_y^i)$ of the ground state can be easily determined, in consistency with the ED results in Figs.~\ref{energy}(a) and~\ref{energy}(b). Meanwhile, we calculate the charge-hole gap in the charge channel $\Delta_q=(E_0(N_{\uparrow}+1,N_{\downarrow})+E_0(N_{\uparrow}-1,N_{\downarrow})-2E_0(N_{\uparrow},N_{\downarrow}))/2$, and the magnon spin gap in the spin channel $\Delta_s=E_0(N_{\uparrow}+1,N_{\downarrow}-1)-E_0(N_{\uparrow},N_{\downarrow})$ for different system sizes. As shown in Figs.~\ref{gap}(a) and~\ref{gap}(b) for different system sizes, the finite size scaling of $\Delta_q$ remains a large finite value, which serves as a primary signature of an incompressible charge Hall phase, while the finite size scaling of $\Delta_s$ goes to a vanishing small value in the thermodynamic limit, signaling a compressible spin superfluid.

Next we extract the Chern number matrix $\mathbf{C}=\begin{pmatrix}
C_{\uparrow\uparrow} & C_{\uparrow\downarrow} \\
C_{\downarrow\uparrow} & C_{\downarrow\downarrow} \\
\end{pmatrix}$ for a two-component system, related to the Hall conductance~\cite{Niu1985,Sheng2003,Sheng2006}. With twisted boundary conditions $\psi(\rr_{\sigma}+N_{\alpha})=\psi(\rr_{\sigma})\exp(i\theta_{\sigma}^{\alpha})$ where $\theta_{\sigma}^{\alpha}$ is the twisted angle for spin $\sigma$ particles in the $\alpha$ direction, the Chern number of the many-body ground state wavefunction $\psi$ is defined in the parameter plane $(\theta_{\sigma}^{x},\theta_{\sigma'}^{y})$ as
\begin{align}
  C_{\sigma,\sigma'}=\int\int \frac{d\theta_{\sigma}^{x}d\theta_{\sigma'}^{y}}{2\pi}F^{xy}(\theta_{\sigma}^{x},\theta_{\sigma'}^{y}),\nonumber
\end{align}
where $F^{xy}=\mathbf{Im}(\langle{\frac{\partial\psi}{\partial\theta_{\sigma}^x}}|{\frac{\partial\psi}{\partial\theta_{\sigma'}^y}}\rangle
-\langle{\frac{\partial\psi}{\partial\theta_{\sigma'}^y}}|{\frac{\partial\psi}{\partial\theta_{\sigma}^x}}\rangle)$ is the Berry curvature and the off-diagonal part $C_{\uparrow\downarrow}$ is related to the drag Hall conductance between spin $\uparrow$ and spin $\downarrow$. With $\theta_{\uparrow}^{x}=\theta_{\downarrow}^{x}=\theta^{x}$ and $\theta_{\uparrow}^{y}=\theta_{\downarrow}^{y}=\theta^{y}$, the charge Chern number, related to charge Hall conductance, reads $C_{q}=\sum_{\sigma,\sigma'}C_{\sigma,\sigma'}=\nu$.
Similarly, with $\theta_{\uparrow}^{x}=-\theta_{\downarrow}^{x}=\theta^{x}$ and $\theta_{\uparrow}^{y}=-\theta_{\downarrow}^{y}=\theta^{y}$, we can also define the spin Chern number of the many-body ground state wavefunction, related to spin Hall conductance, as $C_{s}=C_{\uparrow,\uparrow}+C_{\downarrow,\downarrow}-C_{\uparrow,\downarrow}-C_{\downarrow,\uparrow}$.

\begin{figure}[t]
  \includegraphics[height=1.8in,width=3.4in]{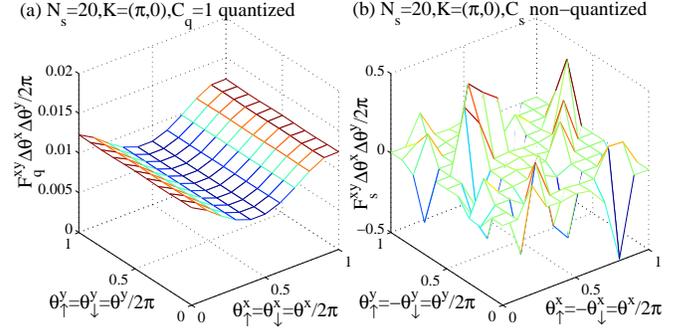}
  \caption{\label{flux} (Color online) Numerical ED results for Berry curvatures $F^{xy}\Delta\theta_{\sigma}^{x}\Delta\theta_{\sigma'}^{y}/2\pi$ of the $K=(\pi,0)$ ground state of two-component fermionic systems $N_{\uparrow}=N_{\downarrow}=5,N_s=2\times2\times N_{\uparrow}$ at $U\gg t,V=0$ on the checkerboard lattice in the parameter plane: (a) $(\theta_{\uparrow}^{x}=\theta_{\downarrow}^{x}=\theta^{x},\theta_{\uparrow}^{y}=\theta_{\downarrow}^{y}=\theta^{y})$ and (b) $(\theta_{\uparrow}^{x}=-\theta_{\downarrow}^{x}=\theta^{x},\theta_{\uparrow}^{y}=-\theta_{\downarrow}^{y}=\theta^{y})$.}
\end{figure}

For the single ground state of two-component fermions, by numerically calculating the Berry curvatures using $m\times m$ mesh squares in the boundary phase space with $m\geq10$ we obtain the quantized topological invariant $C_{q}=1$ with a well-defined smooth Berry curvature, and the non-quantized topological invariant $C_{s}$ with strongly fluctuating Berry curvature, as indicated in Figs.~\ref{flux}(a) and~\ref{flux}(b) respectively, consistent with the gapped charge insulation and gapless spin superfluidity of the bulk.

\begin{figure}[t]
  \includegraphics[height=1.8in,width=3.4in]{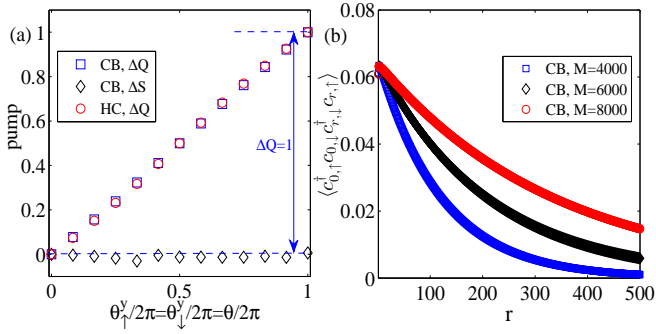}
  \caption{\label{pump} (Color online) (a) The charge and spin transfer for two-component fermions on the $N_y=4$ cylinder at $\nu=1,U\gg t,V=0$ with inserting flux $\theta_{\uparrow}^{y}=\theta_{\downarrow}^{y}=\theta$ for both topological checkerboard and honeycomb lattices. (b) The off-diagonal long range order of the particle-hole pair $\langle c_{0,\uparrow}^{\dag}c_{0,\downarrow}c_{r,\downarrow}^{\dag}c_{r,\uparrow}\rangle$ versus the lattice distance $r$ along the $x$ direction. As the maximal bond dimension is increased, $\langle c_{0,\uparrow}^{\dag}c_{0,\downarrow}c_{r,\downarrow}^{\dag}c_{r,\uparrow}\rangle$ tends to a finite large value for $r\gg1$ when the DMRG results are more and more converged. }
\end{figure}

For larger system sizes, we further calculate the charge pumping under the insertion of flux quantum on infinite cylinder systems in connection to the quantized Hall conductance using DMRG~\cite{Gong2014}. For Halperin (111) state, it is expected that a quantized charge will be pumped from the right side to the left side by inserting a charge flux $\theta_{\uparrow}^{y}=\theta_{\downarrow}^{y}=\theta$ from $\theta=0$ to $\theta=2\pi $. The net transfer of the total charge from the right side to the left side is encoded by $Q(\theta)=N_{\uparrow}^{L}+N_{\downarrow}^{L}=tr[\widehat{\rho}_L(\theta)\widehat{Q}]$ ($\widehat{\rho}_L$ the reduced density matrix of the left part). In order to quantify the spin Hall conductance, we also define the spin transfer $\Delta S$ by $S(\theta)=0.5\times(N_{\uparrow}^{L}-N_{\downarrow}^{L})=tr[\widehat{\rho}_L(\theta)\widehat{S}]$ in analogy to the charge transfer. As shown in Fig.~\ref{pump}(a), a unit charge $\Delta Q=Q(2\pi)-Q(0)\simeq C_{q}=1$ is pumped upon threading one flux quanta with $\theta_{\uparrow}=\theta_{\downarrow}=\theta$ for two-component fermions, without spin pumping $\Delta S=0$, verifying the robustness of quantized charge Hall effect at $\nu=1$.

Moreover, we calculate the off-diagonal long range order $\langle c_{0,\uparrow}^{\dag}c_{0,\downarrow}c_{r,\downarrow}^{\dag}c_{r,\uparrow}\rangle$, describing an exciton condensate as a fermion of spin $\uparrow$ bound to a hole of spin $\downarrow$ forming an exciton pair. The non-vanishing finite value of $\langle c_{0,\uparrow}^{\dag}c_{0,\downarrow}c_{r,\downarrow}^{\dag}c_{r,\uparrow}\rangle$ in the long distance $r\gg1$ is the hallmark signature of the emergence of Bose-Einstein condensation of excitons with symmetry-broken order parameter $\langle\psi|c_{r,\uparrow}^{\dag}c_{r,\downarrow}|\psi\rangle$ from the variational wavefunction in Eq.~\ref{wave}, as indicated in Fig.~\ref{pump}(b). Alternatively, in terms of the transverse spin-flip operator $S^{+}_r\propto c_{r,\uparrow}^{\dag}c_{r,\downarrow}$, it indeed describes the long-range order of quantum Hall ferromagnetism in the $XY$ easy-plane. At fractional fillings, another distinct Ising type of quantum Hall ferromagnets is discussed in Chern bands with high Chern number~\cite{Kumar2014}.

\begin{figure}[t]
  \includegraphics[height=1.95in,width=3.4in]{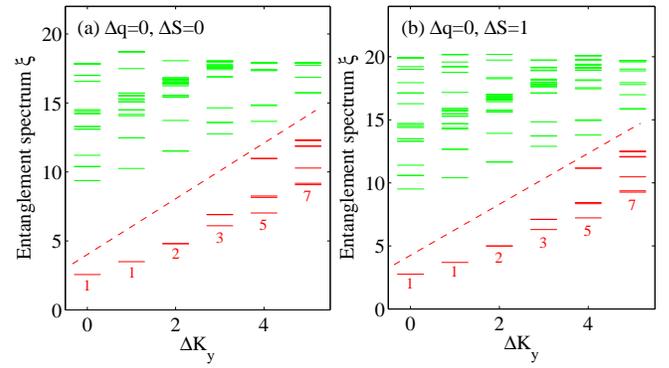}
  \caption{\label{es2} (Color online) Chiral edge mode identified from the momentum-resolved entanglement spectrum for two-component fermions on the $N_y=6$ cylinder at $\nu=1,U\gg t,V=0$. The horizontal axis shows the relative momentum $\Delta K=K_y-K_{y}^{0}$ (in units of $2\pi/N_y$). The numbers below the red dashed line label the nearly degenerating pattern with different momenta: $1,1,2,3,5,7,\cdots$ for different charge and spin sectors (a) $\Delta q=0,\Delta S=0$ and (b) $\Delta q=0,\Delta S=1$.}
\end{figure}

Another ``fingerprint'' of Halperin (111) state is the characteristic chiral edge mode which can be revealed through the low-lying entanglement spectrum in the bulk~\cite{Li2008}. Different from usual two-component quantum Hall states, one of the eigenvalues of $\mathbf{K}=\begin{pmatrix}
1 & 1\\
1 & 1\\
\end{pmatrix}$ is positive while the other is zero, that is, Halperin (111) state hosts only one chiral branch of edge mode~\cite{Wen1992edge,Wen1995}. Here we examine the structure of momentum-resolved entanglement spectrum for different cylinder widths $N_y=4,6$. On the $N_y=4$ cylinder, we observe only one forward-moving branch of low-lying bulk entanglement spectrum with the level counting $1,1,2,3$ for different charge and spin sectors. Similarly, as shown in Figs.~\ref{es2}(a) and~\ref{es2}(b), only one forward-moving branch of low-lying bulk entanglement spectrum with the level counting $1,1,2,3,5,7$ is obtained for different charge sectors on the $N_y=6$ cylinder. Nevertheless, this level counting is consistent with $SU(2)_1$ Wess-Zumino-Witten conformal field theory with the central charge $c=1$, implying the gapless nature of chiral edge modes.

\begin{figure}[b]
  \includegraphics[height=1.8in,width=3.4in]{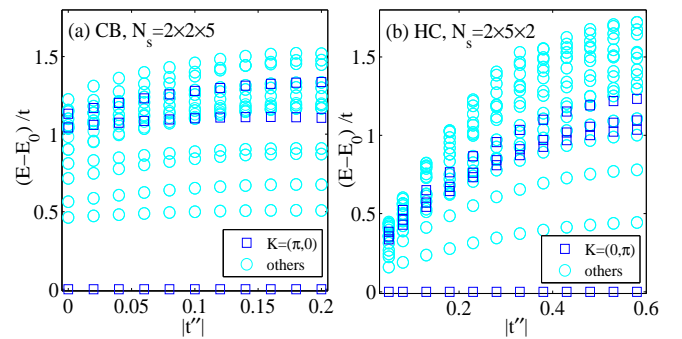}
  \caption{\label{stability} (Color online) Numerical ED results for the low energy evolution of two-component fermionic systems as a function of $t''$ in different topological lattices at $\nu=1,U\gg t,V=0$ for (a) $\pi$-flux checkerboard lattice and (b) Haldane-honeycomb lattice.}
\end{figure}

In view of the current Haldane-honeycomb experiment where $t''$ is negligible and the topological band becomes significantly dispersive~\cite{Jotzu2014}, it is natural and important to take into account the effect of dispersive band structures controlled by the weak next-next-nearest-neighbour tunnel coupling $t''$, on the stability of this exciton condensate under strong Hubbard repulsion. As shown in Figs.~\ref{stability}(a) and~\ref{stability}(b), when $t''$ is tuned down away from the flat band limit, we observe that the unique ground state evolves smoothly and persists with a moderately large protecting energy gap of the order $0.1t$ even in the weakly tunneling regime $|t''|\ll t,t'$ at $\nu=1,U\gg t$. The robustness of Halperin (111) state, regardless of the longer range hopping or band dispersion in different topological lattices, makes it very promising and straightforward to be detected in experiments through the Hall drag transport or the transverse drifting motion of the center-of-mass under external longitudinal field gradient~\cite{Aidelsburger2015}.

\subsection{Halperin (333) state}\label{H333}

\begin{figure}[t]
  \includegraphics[height=1.8in,width=3.4in]{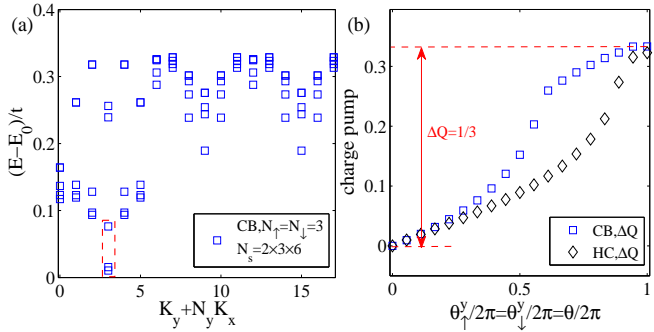}
  \caption{\label{en333} (Color online) (a) Numerical ED results for the low energy spectrum of two-component fermionic systems in topological checkerboard lattice at $\nu=1/3,U,V\gg t$. The red dashed box depicts three-fold quasi-degeneracy. (b) The charge transfer for two-component fermions on the $N_y=3$ cylinder at $\nu=1/3,U,V\gg t$ with inserting flux $\theta_{\uparrow}^{y}=\theta_{\downarrow}^{y}=\theta$ for both topological checkerboard and honeycomb lattices.}
\end{figure}

We now turn to analyze the possible emergence of Halperin fractional quantum Hall state characterized by the $\mathbf{K}=\begin{pmatrix}
3 & 3\\
3 & 3\\
\end{pmatrix}$ matrix at $\nu=1/3$ under both strong onsite and nearest-neighbor Hubbard repulsions $U,V\gg t$, hosting fractionalized quasiparticles in analogy to the Laughlin fractional quantum Hall state. For small finite system sizes $N_s=12,18$, our ED calculations gives three-fold quasi-degenerate ground states at the total momentum $K_y+N_yK_x$ given by the generalized Pauli exclusion principle~\cite{Regnault2011}, i.~e. no more than one particle is allowed occupy within any consecutive three orbitals in the lowest Chern band, as indicated in Fig.~\ref{en333}(a). We numerically confirm that under the insertion of three flux quanta, the system returns to itself, and the ground state hosts a fractionally quantized Chern number $C_q=1/3$ in the parameter plane $(\theta_{\uparrow}^{x}=\theta_{\downarrow}^{x},\theta_{\uparrow}^{y}=\theta_{\downarrow}^{y})$, demonstrating its $1/3$ fractional quantization of quasiparticles. For larger system sizes, as indicated in Fig.~\ref{en333}(b), our DMRG simulation of the total charge pumping under the insertion of flux quantum $\theta_{\uparrow}^{y}=\theta_{\downarrow}^{y}$ on infinite cylinder systems gives a well-quantized value $\Delta Q\simeq1/3=C_q$, consistent with the analysis of ED study.

\section{Conclusion}\label{summary}

In summary, we have proved numerically that two-component fermions in topological lattice models could realize Halperin $(mmm)$ quantum Hall states at commensurate partial fillings $\nu=1/m$ (odd $m=1,3$) in the lowest Chern band, with topological properties characterized by the $\mathbf{K}=\begin{pmatrix}
m & m\\
m & m\\
\end{pmatrix}$ matrix. For Halperin $(111)$ state, we demonstrate that it is an intercomponent exciton condensate of particle and hole pairs bound by the effective attractive interaction between particles and holes, when the onsite Hubbard interaction between intercomponent particles is repulsive, along with integer quantized Hall conductance and one chiral edge mode. For Halperin $(333)$ state, we qualitatively identify its fractionally quantized topological nature from the degenerate ground state manifold and one-third quantized Hall conductance, similar to Laughlin $\nu=1/3$ fractional quantum Hall effect. At experimental side, our two-component flat band models would be paradigmatic examples of a Hamiltonian featuring topological exciton condensation purely driven by local interaction, which is of sufficient feasibility to be realized for current optical Haldane-honeycomb lattice experiments in cold atoms. We believe that this work would offer an alternative route for the study of exotic topological excitonic insulator on designed band structures~\cite{Seradjeh2009,Hu2018}, and excite a more extensive investigation of the fate of topological exciton condensate in many other topological band systems, such as moir\'e exciton~\cite{Wu2017,Tran2019,Jin2019} in twisted multilayer graphene when strong electronic correlation is introduced.

\begin{acknowledgements}
This work is supported by start-up funding from Westlake University, and the NSFC under Grant No. 11974288 (T.S.Z., W.Z.). T.S.Z. also acknowledges the support from the Natural Science Foundation of Zhejiang Province of China under Grant No. LQ20A040001. D.N.S is supported by National Science Foundation PREM under Grant No. DMR-1828019.
\end{acknowledgements}


\begin{thebibliography}{99}
\bibitem{Halperin1983} B.~I. Halperin, {\it Theory of the quantized Hall conductance}, Helv. Phys. Acta, {\bf 56}, 75 (1983).
\bibitem{Wen1992a} X.~G. Wen and A. Zee, {\it Classification of Abelian quantum Hall states and matrix formulation of topological fluids}, Phys. Rev. B {\bf 46}, 2290 (1992).
\bibitem{Wen1992b} X.~G. Wen and A. Zee, {\it Shift and spin vector: New topological quantum numbers for the Hall fluids}, Phys. Rev. Lett. {\bf 69}, 953 (1992).
\bibitem{Blok1990a} B. Blok and X.~G. Wen, {\it Effective theories of the fractional quantum Hall effect at generic filling fractions}, Phys. Rev. B {\bf 42}, 8133 (1990).
\bibitem{Blok1990b} B. Blok and X.~G. Wen, {\it Effective theories of the fractional quantum Hall effect: Hierarchy construction}, Phys. Rev. B {\bf 42}, 8145 (1990).
\bibitem{Blok1991} B. Blok and X. G. Wen, {\it Structure of the microscopic theory of the hierarchical fractional quantum Hall effect}, Phys. Rev. B {\bf 43}, 8337 (1991).
\bibitem{Eisenstein1992} J.~P. Eisenstein, G.~S. Boebinger, L.~N. Pfeiffer, K.~W. West, and S. He, {\it New fractional quantum Hall state in double-layer two-dimensional electron systems}, Phys. Rev. Lett. {\bf 68}, 1383 (1992).
\bibitem{Suen1992} Y.~W. Suen, L.~W. Engel, M.~B. Santos, M. Shayegan, and D.~C. Tsui, {it Observation of a \ensuremath{\nu}=1/2 fractional quantum Hall state in a double-layer electron system}, Phys. Rev. Lett. {\bf 68}, 1379 (1992).
\bibitem{Wen1992} X.-G. Wen and A. Zee, {\it Neutral superfluid modes and ``magnetic'' monopoles in multilayered quantum Hall systems}, Phys. Rev. Lett. {\bf 69}, 1811 (1992).
\bibitem{Moon1995} K. Moon, H. Mori, K. Yang, S.~M. Girvin, A.~H. MacDonald, L. Zheng, D. Yoshioka, and S.-C. Zhang, {\it Spontaneous interlayer coherence in double-layer quantum Hall systems: Charged vortices and Kosterlitz-Thouless phase transitions}, Phys. Rev. B {\bf 51}, 5138 (1995).
\bibitem{Yang1996} K. Yang, K. Moon, Lotfi Belkhir, H. Mori, S.~M. Girvin, A.~H. MacDonald, L. Zheng, and D. Yoshioka, {\it Spontaneous interlayer coherence in double-layer quantum Hall systems: Symmetry-breaking interactions, in-plane fields, and phase solitons}, Phys. Rev. B {\bf 54}, 11644 (1996).
\bibitem{Yoshioka1988} D. Yoshioka, A.~H. MacDonald, and S.~M. Girvin, {\it Connection between spin-singlet and hierarchical wave functions in the fractional quantum Hall effect}, Phys. Rev. B {\bf 38}, 3636(R) (1988).
\bibitem{Yoshioka1989} D. Yoshioka, A.~H. MacDonald, and S.~M. Girvin, {\it Fractional quantum Hall effect in two-layered systems}, Phys. Rev. B {\bf 39}, 1932 (1989).
\bibitem{He1991} S. He, X.~C. Xie, S. Das Sarma, and F.~C. Zhang, {\it Quantum Hall effect in double-quantum-well systems}, Phys. Rev. B {\bf 43}, 9339(R) (1991).
\bibitem{He1993} S. He, S. Das Sarma, and X.~C. Xie, {\it Quantized Hall effect and quantum phase transitions in coupled two-layer electron systems}, Phys. Rev. B {\bf 47}, 4394 (1993).
\bibitem{Zeng2017} T.-S. Zeng, W. Zhu, and D. N. Sheng, {\it Two-component quantum Hall effects in topological flat bands}, Phys. Rev. B {\bf 95}, 125134 (2017).
\bibitem{Zeng2018} T.-S. Zeng and D. N. Sheng, {\it $\mathrm{SU}(N)$ fractional quantum Hall effect in topological flat bands}, Phys. Rev. B {\bf 97}, 035151 (2018).
\bibitem{Zeng2019} T.-S. Zeng, D.~N. Sheng, and W. Zhu, {\it Topological characterization of hierarchical fractional quantum Hall effects in topological flat bands with SU($N$) symmetry}, Phys. Rev. B {\bf 100}, 075106 (2019).
\bibitem{Zeng2020} T.-S. Zeng, D.~N. Sheng, and W. Zhu, {\it Continuous phase transition between bosonic integer quantum Hall liquid and a trivial insulator: Evidence for deconfined quantum criticality}, Phys. Rev. B {\bf 101}, 035138 (2020).
\bibitem{Eisenstein2014} J.~P. Eisenstein, {\it Exciton Condensation in Bilayer Quantum Hall Systems}, Annu. Rev. of Condens. Matter Phys. {\bf 5}, 159 (2014).
\bibitem{Kellogg2002} M. Kellogg, I.~B. Spielman, J.~P. Eisenstein, L.~N. Pfeiffer, and K.~W. West, {\it Observation of Quantized Hall Drag in a Strongly Correlated Bilayer Electron System}, Phys. Rev. Lett. {\bf 88}, 126804 (2002).
\bibitem{Tutuc2004} E. Tutuc, M. Shayegan, and D.~A. Huse, {\it Counterflow Measurements in Strongly Correlated GaAs Hole Bilayers: Evidence for Electron-Hole Pairing}, Phys. Rev. Lett. {\bf 93}, 036802 (2004).
\bibitem{Finck2010} A.~D.~K. Finck, J. P. Eisenstein, L.~N. Pfeiffer, and K.~W. West,  {\it Quantum Hall Exciton Condensation at Full Spin Polarization}, Phys. Rev. Lett. {\bf 104}, 016801 (2010).
\bibitem{Nandi2012} D. Nandi, A.~D.~K. Finck, J.~P. Eisenstein, L.~N. Pfeiffer, and K.~W. West, {\it Exciton condensation and perfect Coulomb drag}, Nature {\bf 488}, 481 (2012).
\bibitem{Liu2017} X. Liu, K. Watanabe, T. Taniguchi, B.~I. Halperin, and P. Kim, {\it Quantum Hall drag of exciton condensate in graphene}, Nature Phys. {\bf 13}, 746 (2017).
\bibitem{Li2017} J.~I.~A. Li, T. Taniguchi, K. Watanabe, J. Hone, and C.~R. Dean, {\it Excitonic superfluid phase in double bilayer graphene}, Nature Phys. {\bf 13}, 751 (2017).
\bibitem{Eisenstein2019} J.~P. Eisenstein, L.~N. Pfeiffer, and K.~W. West, {\it Precursors to Exciton Condensation in Quantum Hall Bilayers}, Phys. Rev. Lett. {\bf 123}, 066802 (2019).
\bibitem{Li2019} J.~I.~A. Li, Q. Shi, Y. Zeng, K. Watanabe, T. Taniguchi, J. Hone, and C.~R. Dean, {\it Pairing states of composite fermions in double-layer graphene}, Nature Phys. {\bf 15}, 898 (2019).
\bibitem{Liu2019} X. Liu, Z. Hao, K. Watanabe, T. Taniguchi, B.~I. Halperin, and P. Kim, {\it Interlayer fractional quantum Hall effect in a coupled graphene double layer}, Nature Phys. {\bf 15}, 893 (2019).
\bibitem{Sun2011} K. Sun, Z. Gu, H. Katsura, and S. Das Sarma, {\it Nearly Flatbands with Nontrivial Topology}, Phys. Rev. Lett. {\bf 106}, 236803 (2011).
\bibitem{Neupert2011} T. Neupert, L. Santos, C. Chamon, and C. Mudry, {\it Fractional Quantum Hall States at Zero Magnetic Field}, Phys. Rev. Lett. {\bf 106}, 236804 (2011).
\bibitem{Sheng2011} D.~N. Sheng, Z. Gu, K. Sun, and L. Sheng, {\it Fractional quantum Hall effect in the absence of Landau levels}, Nat. Commun. {\bf 2}, 389 (2011).
\bibitem{Tang2011} E. Tang, J.-W. Mei, and X.-G. Wen, {\it High-Temperature Fractional Quantum Hall States}, Phys. Rev. Lett. {\bf 106}, 236802 (2011).
\bibitem{Wang2011} Y.-F. Wang, Z.-C. Gu, C.-D. Gong, and D.~N. Sheng, {\it Fractional Quantum Hall Effect of Hard-Core Bosons in Topological Flat Bands}, Phys. Rev. Lett. {\bf 107}, 146803 (2011).
\bibitem{Regnault2011} N. Regnault and B.~A. Bernevig, {\it Fractional Chern Insulator}, Phys. Rev. X {\bf 1}, 021014 (2011).
\bibitem{BL2013} E.~J. Bergholtz and Z. Liu, {\it Topological flat band models and fractional Chern insulators}, Int. J. Mod. Phys. B {\bf 27}, 1330017 (2013).
\bibitem{Cayssol2013} J. Cayssol, B. D\'{o}ra, F. Simon, and R. Moessner, {\it Floquet topological insulators}, Phys. Status Solidi RRL {\bf 7}, 101 (2013).
\bibitem{Jotzu2014} G. Jotzu, M. Messer, R. Desbuquois, M. Lebrat, T. Uehlinger, D. Greif and T. Esslinger, {\it Experimental realization of the topological Haldane model with ultracold fermions}, Nature {\bf 515}, 237 (2014).
\bibitem{Koepsell2020} J. Koepsell, S. Hirthe, D. Bourgund, P. Sompet, J. Vijayan, G. Salomon, C. Gross, and I. Bloch, {\it Robust Bilayer Charge-Pumping for Spin- and Density-Resolved Quantum Gas Microscopy}, \href{https://arxiv.org/abs/2002.07577}{arXiv:2002.07577.}
\bibitem{Zhang2019} Y.-H. Zhang, D. Mao, Y. Cao, P. Jarillo-Herrero, and T. Senthil, {\it Nearly flat Chern bands in moir\'e superlattices}, Phys. Rev. B {\bf 99}, 075127 (2019).
\bibitem{Lee2019} J. Y. Lee, E. Khalaf, S. Liu, X. Liu, Z. Hao, P. Kim, and A. Vishwanath, {\it Theory of correlated insulating behaviour and spin-triplet superconductivity in twisted double bilayer graphene}, Nat. Commun. {\bf 10}, 5333 (2019).
\bibitem{Dai2019} J. Liu, Z. Ma, J. Gao, X. Dai, {\it Quantum Valley Hall Effect, Orbital Magnetism, and Anomalous Hall Effect in Twisted Multilayer Graphene Systems}, Phys. Rev. X {\bf 9}, 031021 (2019).
\bibitem{Chen2019} G. Chen, L. Jiang, S. Wu, B. Lyu, H. Li, B.~L. Chittari, K. Watanabe, T. Taniguchi, Z. Shi, J. Jung, {\it et al.}, {\it Evidence of a gate-tunable Mott insulator in a trilayer graphene moir\'e superlattice}, Nature Phys. {\bf 15}, 237 (2019).
\bibitem{Chen2020} G. Chen, A.~L. Sharpe, E.~J. Fox, Y.-H. Zhang, S. Wang, L. Jiang, B. Lyu, H. Li, K. Watanabe, T. Taniguchi, {\it et al.}, {\it Tunable correlated Chern insulator and ferromagnetism in a moir\'e superlattice}, Nature {\bf 579}, 56 (2020).
\bibitem{Spanton2017} E.~M. Spanton, A.~A. Zibrov, H. Zhou, T. Taniguchi, K. Watanabe, M.~P. Zaletel, A.~F. Young, {\it Observation of fractional Chern insulators in a van der Waals heterostructure}, Science {\bf 360}, 62 (2018).
\bibitem{Zeng2015} T.-S. Zeng and L. Yin, {\it Fractional quantum Hall states of dipolar gases in Chern bands}, Phys. Rev. B {\bf 91}, 075102 (2015).
\bibitem{Wen1995} X.-G. Wen, {\it Topological orders and edge excitations in fractional quantum Hall states}, Adv. Phys. {\bf 44}, 405 (1995).
\bibitem{Niu1985} Q. Niu, D.~J. Thouless, and Y.-S. Wu, {\it Quantized Hall conductance as a topological invariant}, Phys. Rev. B {\bf 31}, 3372 (1985).
\bibitem{Sheng2003} D.~N. Sheng, L. Balents, and Z. Wang, {\it Phase Diagram for Quantum Hall Bilayers at $\ensuremath{\nu}=1$}, Phys. Rev. Lett. {\bf 91}, 116802 (2003).
\bibitem{Sheng2006} D.~N. Sheng, Z.-Y. Weng, L. Sheng, and F.~D.~M. Haldane, {\it Quantum Spin-Hall Effect and Topologically Invariant Chern Numbers}, Phys. Rev. Lett. {\bf 97}, 036808 (2006).
\bibitem{Gong2014} S. S. Gong, W. Zhu, and D. N. Sheng, {\it Emergent Chiral Spin Liquid: Fractional Quantum Hall Effect in a Kagome Heisenberg Model}, Sci. Rep. {\bf 4}, 6317 (2014).
\bibitem{Kumar2014} A. Kumar, R. Roy, and S.~L. Sondhi, {\it Generalizing quantum Hall ferromagnetism to fractional Chern bands}, Phys. Rev. B {\bf 90}, 245106 (2014).
\bibitem{Li2008} H. Li and F.~D.~M. Haldane, {\it Entanglement Spectrum as a Generalization of Entanglement Entropy: Identification of Topological Order in Non-Abelian Fractional Quantum Hall Effect States}, Phys. Rev. Lett. {\bf 101}, 010504 (2008).
\bibitem{Wen1992edge} X.-G. Wen, {\it Theory of the edge states in fractional quantum Hall effects}, Int. J. Mod. Phys. B {\bf 6}, 1711 (1992).
\bibitem{Aidelsburger2015} M. Aidelsburger, M. Lohse, C. Schweizer, M. Atala, J.~T. Barreiro, S. Nascimb\`{e}ne, N.~R. Cooper, I. Bloch, and N. Goldman, {\it Measuring the Chern number of Hofstadter bands with ultracold bosonic atoms}, Nature Phys. {\bf 11}, 162 (2015).
\bibitem{Seradjeh2009} B. Seradjeh, J. E. Moore, and M. Franz, {\it Exciton Condensation and Charge Fractionalization in a Topological Insulator Film}, Phys. Rev. Lett. {\bf 103}, 066402 (2009).
\bibitem{Hu2018} Y. Hu, J.~W.~F. Venderbos, and C.~L. Kane, {\it Fractional Excitonic Insulator}, Phys. Rev. Lett. {\bf 121}, 126601 (2018).
\bibitem{Wu2017} F. Wu, T. Lovorn, and A.~H. MacDonald, {\it Topological Exciton Bands in Moir\'e Heterojunctions}, Phys. Rev. Lett. {\bf 118}, 147401 (2017).
\bibitem{Tran2019} K. Tran, G. Moody, F. Wu, X. Lu, J. Choi, K. Kim, A. Rai, D.~A. Sanchez, J. Quan, A. Singh, {\it et al.}, {\it Evidence for moir\'e excitons in van der Waals heterostructures}, Nature {\bf 567}, 71 (2019).
\bibitem{Jin2019} C. Jin, E.~C. Regan, A. Yan, M. Iqbal Bakti Utama, D. Wang, S. Zhao, Y. Qin, S. Yang, Z. Zheng, S. Shi, {\it et al.}, {\it Observation of moir\'e excitons in WSe$_2$/WS$_2$ heterostructure superlattices}, Nature {\bf 567}, 76 (2019).
\end{thebibliography}
\end{document}